\documentclass[sn-nature,iicol,Numbered]{sn-jnl}
\usepackage{graphicx}%
\usepackage{multirow}%
\usepackage{amsmath,amssymb,amsfonts,bbm}%
\usepackage{mathbbol}
\usepackage{amsthm}%
\usepackage{mathrsfs}%
\usepackage[title]{appendix}%
\usepackage{xcolor}%
\usepackage{textcomp}%
\usepackage{manyfoot}%
\usepackage{booktabs}%
\usepackage{algorithm}%
\usepackage{algorithmicx}%
\usepackage{algpseudocode}%
\usepackage{listings}%
\usepackage{cleveref}
\usepackage{bm}
\usepackage{pdfpages}
\usepackage{float}
\graphicspath{{./figures/}}

\theoremstyle{thmstyleone}%
\theoremstyle{thmstyletwo}%

\theoremstyle{thmstylethree}%

\usepackage[normalem]{ulem}  % \sout{old text} for strikeout

\renewcommand{\sout}{\bgroup \color{red} \ULdepth=-.5ex \ULset}

\usepackage{tikz,xcolor,hyperref}
\definecolor{lime}{HTML}{A6CE39}
\DeclareRobustCommand{\orcidicon}{
	\begin{tikzpicture}
	\draw[lime, fill=lime] (0,0) 
	circle [radius=0.16] 
	node[white] {{\fontfamily{qag}\selectfont \tiny ID}};
	\draw[white, fill=white] (-0.0625,0.095) 
	circle [radius=0.007];
	\end{tikzpicture}
	\hspace{-2mm}
}

\foreach \x in {A, ..., Z}{%
	\expandafter\xdef\csname orcid\x\endcsname{\noexpand\href{https://orcid.org/\csname orcidauthor\x\endcsname}{\noexpand\orcidicon}}
}

\def\Eq#1{Eq.~\labelcref{#1}}
\def\Fig#1{Fig.~\labelcref{#1}}

\allowdisplaybreaks[1]
\begin{document}

\title[Article Title]{Universality of pseudo-Goldstone damping near critical points}

\author[1]{Yang-yang Tan}
\equalcont{These authors contributed equally to this work.}

\author[1]{Yong-rui Chen}
\equalcont{These authors contributed equally to this work.}

\author*[1,2]{Wei-jie Fu}
\email{wjfu@dlut.edu.cn}

\author*[1]{Wei-Jia Li}
\email{weijiali@dlut.edu.cn}
 
\affil[1]{School of Physics, Dalian University of Technology, Dalian, 116024, China}	
\affil[2]{Shanghai Research Center for Theoretical Nuclear Physics, NSFC and Fudan University, Shanghai 200438, China}

\abstract{Real-time dynamics of strongly correlated systems, in particular its critical dynamics near phase transitions, have been always on the cutting edge of studies in diverse fields of physics, e.g., high energy physics, condensed matter, holography, etc. In this work, we investigate the critical damping of collective modes associated with spontaneous breaking of approximate symmetries, which are called pseudo-Goldstone modes, in strongly correlated systems. Using the Schwinger-Keldysh field theory, we find a universal pseudo-Goldstone damping via the critical O($N$) model that has never been found before by other approaches. Different from the conventional damping found in holography and hydrodynamics, the new one is controlled by critical fluctuations, hence is invisible in mean-field systems or strongly correlated systems with classical gravity duals. Since the critical damping depends solely on the universalities of the critical point, irrespective of the microscopic details, our conclusion should be applicable to a wide class of interacting systems.}
 
% \keywords{(anti)nuclei, nucleosynthesis,  high-energy nuclear collisions, statistical hadronization}

\maketitle

\section{Introduction}\label{sec1}

The notion of spontaneous symmetry breaking (SSB) plays a central role in modern physics. As is well known that when the broken symmetry is global and continuous, a gapless excitation called the Goldstone mode emerges and governs the long wavelength behavior of the system. Usually, symmetries are not exact. In these cases, the would-be gapless excitations acquire a slight mass gap which are often named the pseudo-Goldstone modes. These soft modes are ubiquitous in nature, ranging from the high-energy QCD matter to low energy condensed matter systems \cite{Son:2001ff, Son:2002ci, Gruner:1994zz, Fogler:2000, Fu:2019hdw, Gao:2020fbl, Dupuis:2020fhh, Gunkel:2021oya, Fu:2022gou, Braun:2023qak}.

\begin{figure}[t]   \includegraphics[width=0.40\textwidth]{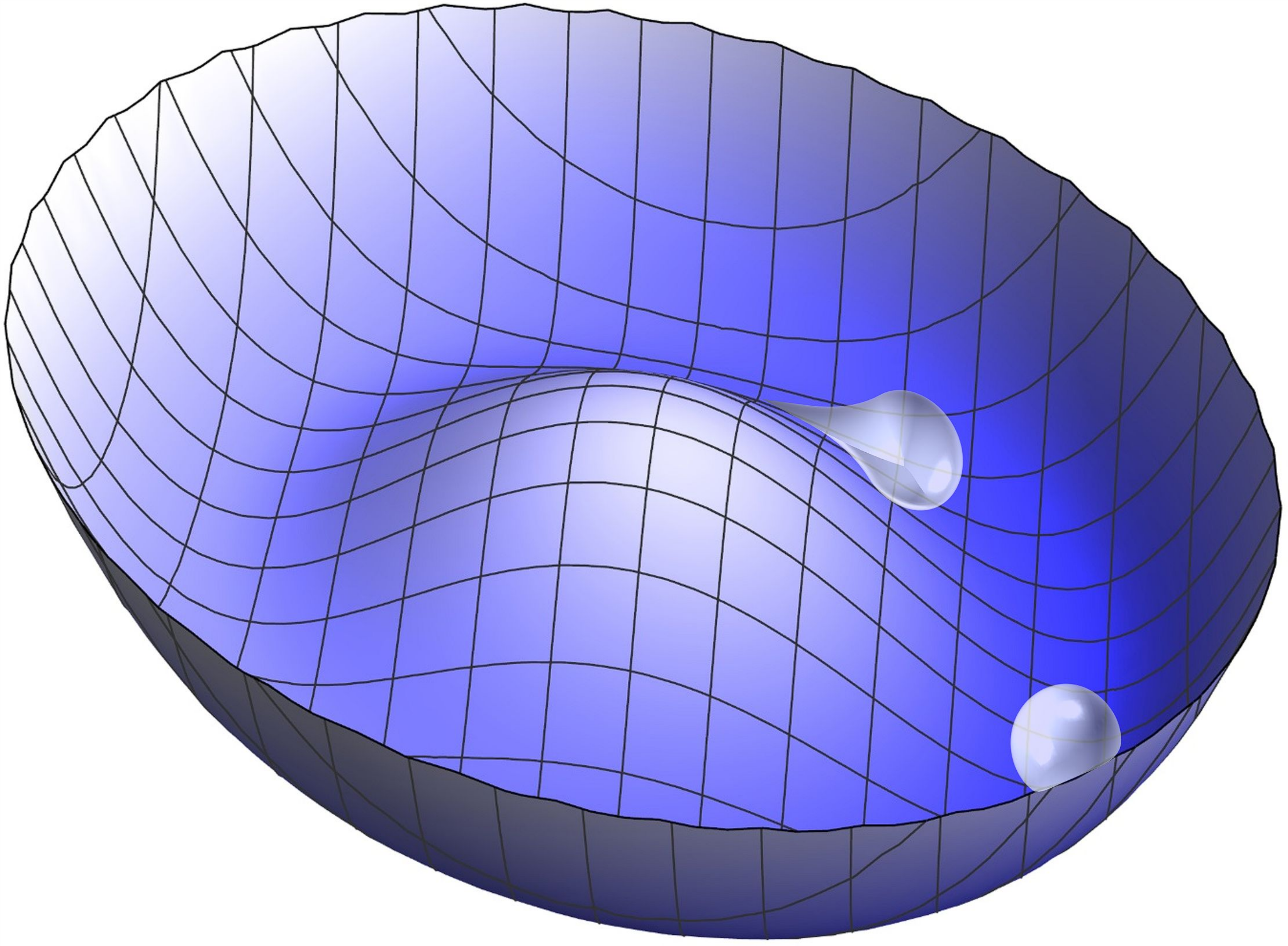}
  \caption{An illustration for the spontaneous breaking of an approximate symmetry and the pseudo-Goldstone modes.  The pseudo-Goldstone modes and their damping are visualized by the droplets and the stretching effect of the droplets.}\label{fig:mexican}
\end{figure}

Over the years, the dynamic properties of pseudo-Goldstone modes have attracted a lot of attention not only because they are crucial for deepening our understanding of real-world systems, but also because they share remarkably universal features that do not depend on microscopic details or even types of broken approximate symmetries, e.g., the `screening mass' of pseudo-Goldstone fields $m_\varphi$ satisfying the celebrated Gell-Mann--Oakes--Renner (GMOR) relation when the system is not very close to the critical point \cite{Gell-Mann:1968hlm,Amoretti:2018tzw, Andrade:2018gqk, Ammon:2019wci, Donos:2021pkk, Ammon:2021pyz, Zhong:2022mok, Cao:2022csq,Argurio:2015wgr, Amoretti:2016bxs}. At finite temperatures, the real part of their dispersion relations, which represents the propagation sector of the real-time response, can be entirely determined by several thermodynamic quantities encoded in the static correlator \cite{Son:2001ff, Son:2002ci}. More recently, it was also noticed in the holographic studies that the damping rate of pseudo-Goldstone modes, $\Omega_\varphi$, which is the imaginary part of their dispersion relation at zero wavenumber is related to the screening mass as well as the transport coefficient in a universal form as follows~\cite{Amoretti:2018tzw,Amoretti:2019cef}
\begin{align}
    \frac{\Omega_\varphi}{m_{\varphi}^2} \simeq &\, D_{\varphi}+\mathcal{O}(m_\varphi^2)\,,\label{eq:universal}
\end{align}
where $D_\varphi$ is the `Goldstone diffusivity' in the purely SSB phase. This relation reveals that pseudo-Goldstone modes can also be damped due to the weak explicit broken symmetry without any topological defects for the spontaneously broken symmetry. This relation was actually noted in the hydrodynamics of QCD based on the argument of positivity of entropy production \cite{Grossi:2020ezz}. Its validity was later on verified also for global $\text{U(1)}$ and $\text{SU(2)}_{\text{L}}\times \text{SU(2)}_{\text{R}}$ symmetries in the context of holography \cite{Ammon:2021pyz,Cao:2022csq} as well as in the effective field theory of quasi-crystals \cite{Baggioli:2020nay,Baggioli:2020haa}. In addition, a more general analysis on the universality of \Eq{eq:universal} has also been carried out within the standard on-shell hydrodynamic approach either by requiring the locality of equations of motion \cite{Delacretaz:2021qqu} or by imposing the second law of thermodynamics \cite{Armas:2021vku} in the presence of external sources. In \Fig{fig:mexican} we show an illustrative diagram for the pseudo-Goldstone modes and their damping.

Nevertheless, neither the standard hydrodynamic approach nor the holographic method (in the large $N$ limit) endows us with a systematic treatment for the fluctuations. Therefore, \Eq{eq:universal} so far can only be trusted in the region away from the criticality where the non-equilibrium fluctuations are significantly suppressed. In this work, we will investigate the pseudo-Goldstone damping when non-equilibrium fluctuations become particularly pronounced near the critical point in a field theory approach. To overcome the difficulties from the strong correlations as well as the significant fluctuations in the critical region, we adopt the functional renormalization group (fRG) within the Schwinger-Keldysh (SK) formalism \cite{Berges:2008sr, Gasenzer:2007za, Pawlowski:2015mia, Corell:2019jxh, Huelsmann:2020xcy, Tan:2021zid, Roth:2021nrd, Braun:2022mgx, Chen:2023tqc, Roth:2023wbp, Batini:2023nan}, which is a powerful nonperturbative method for continuum field theory at finite temperatures. In this approach, all quantum and thermal effects are encoded successively with the evolution of the renormalization group (RG) scale. 

After performing the fRG computation on a critical O($N$) model named `Model A' in the Hohenberg-Halperin classification \cite{Hohenberg:1977ym}, it is found that the conventional relation \Eq{eq:universal} is satisfied only when $T$ is not so close to $T_c$. However, it is not obeyed for any finite $N$ and finite $m_\varphi$  in the vicinity of the critical point where the mass of Higgs mode $m_\sigma\sim m_\varphi$. More specifically, the ratio
$\Omega_\varphi/m_\varphi^2$ displays a novel universal scaling behavior
\begin{equation}
    \frac{\Omega_{\varphi}}{m_\varphi^2}\propto \, m_\varphi^{\Delta_\eta}\,,
    \label{eq:scaling}
\end{equation}
in the critical region that is controlled by $\Delta_\eta\equiv\eta_t-\eta$, the difference between the dynamic anomalous exponent $\eta_t$ and static anomalous exponent $\eta$ due to the large fluctuations. Furthermore, we investigate the dependence of such a scaling regime on the value of $N$, and find that this regime fades away in the mean-field approximation which requires that $N\rightarrow \infty$. This implies that the new behavior of the pseudo-Goldstone damping \Eq{eq:scaling} is invisible in any mean-field systems or strongly-coupled systems with classical gravity duals.

\section{Results}\label{sec2}
\bmhead{A relaxational critical O($N$) model} 
We consider the critical dynamics of a $N$-component real order parameter $\phi_a(t,\bm{x})$ with $a=0, 1, \cdots N-1$  at finite temperature, whose quantum effective action on the Schwinger-Keldysh contour is given by
\begin{align}    
    &\Gamma[\phi_c,\phi_q]\nonumber\\[2ex]
   =&\int\,d^4x\,\Big(Z^{(t)}_{a}\,\phi_{a,q}\, \partial_t\,\phi_{a,c}-Z_{a}^{(i)}\,\phi_{a,q}\, \partial_i^2\,\phi_{a,c}\nonumber\\[2ex]
   &\hspace{-0.3cm}+ V'(\rho_c)\,\phi_{a,q}\,\phi_{a,c}-2 \,Z^{(t)}_{a}\, T\,\phi_{a,q}^2- \sqrt{2} c\,\sigma_q\Big)\,.\label{eq:action}
\end{align}
This is a purely dissipative relaxation model classified as Model A in the seminal paper \cite{Hohenberg:1977ym}. The subscripts `$q$' and `$c$' denote the `quantum' and `classical' fields; $V(\rho_c)$ is the effective potential preserving the O($N$) symmetry with $\rho_c\equiv \phi_c^2/4$ and $V'(\rho_c)$ denotes its derivative; the quadratic term of $\phi_{a,q}$ whose coefficient is fixed by the fluctuation-dissipation theorem creates a Gaussian white noise with temperature $T$. The last term linearly depending on $\sigma_q\equiv\phi_{0,q}$  plays the role of breaking the O($N$) symmetry explicitly with the constant $c$ parametrizing the strength of symmetry breaking. In addition, $Z^{(t,i)}_a$ are wave functions of the temporal and spatial components, both of which depend on $T$ and $c$. Note that in \Eq{eq:action} the RG scale $k$-dependence of the potential and wave functions is ignored. More details about the setup of the model in \Eq{eq:action} within the fRG can be found in Sec.~Methods. When $c=0$, this model allows us to reduce the symmetry from O$(N)$ to O$(N-1)$ spontaneously for a potential $V(\rho_c)$ with the shape of a `Mexican hat' as shown in \Fig{fig:mexican}, such that the fields take the expectation values  $\bar{\phi}_{0,c}\equiv \bar{\sigma}_{c}\neq 0$ and $\bar{\phi}_{i,c}=0$ ($i=1, 2, \cdots N-1$) in the ground state. The expectation values of quantum fields are always vanishing, i.e., $\bar{\phi}_{a,q}=0$. In the broken pattern, the fluctuating fields $\delta\sigma\equiv\delta\phi_0$ and $\varphi_i\equiv\delta\phi_{i}$ play roles of the Higgs mode and the Goldstone modes, respectively. Since the $N-1$ Goldstones are identical in this model, we will omit the subscript `$i$' hereafter. In the following, we will treat O($N$) as an approximate symmetry (i.e., $c$ is non-zero but small) such that the would-be Goldstones acquire a small mass gap.

Here, we should point out that Model A in \Eq{eq:action} neglects the fluctuations of the almost conserved density which in general couples with the pseudo-Goldstone fields. Then, the dynamics is solely controlled by the order parameter, and we will see that the model is not able to describe the propagating feature of the collective modes. However, we here focus only on the dissipative sector that is sufficient for testing the universality of the pseudo-Goldstone damping. For the propagation of the pseudo-Goldstone modes, one can read the extended study of Model G in the Supplementary Information.

All the real-time correlators can be achieved from the effective action in \Eq{eq:action} \cite{Chen:2023tqc}. Dispersion relations of excitations correspond to poles of the retarded correlators which can be computed as follows
\begin{align}
    G_{ab}^{R}=\left( \frac{\delta^2\Gamma[\phi_c,\phi_q]}{\delta \phi_{a,q}\,\delta \phi_{b,c}}\right)^{-1}\,.\label{eq:rtcorrelators}
\end{align}
As a result, this gives rise to the retarded correlator of $\varphi$ fields for small frequency $\omega$ and small wavenumber $q$ in the following form,
\begin{align}
G_{\varphi\varphi}^R(\omega,q)=\frac{1}{-\mathrm{i} Z^{(t)}_\varphi \omega+Z^{(i)}_\varphi \left(q^2+m^2_{\varphi}\right)}\,,\label{eq:retardcorrelators}
\end{align}
with $m^2_{\varphi}\equiv V'(\rho_{c,0})/Z^{(i)}_\varphi$, where $\rho_{c,0}=\rho_{c,0}(c)$ is the value of $\rho_c$ corresponding to the minimum of $V(\rho_c)-c\sqrt{2 \rho_c}$ when $c\neq 0$. Due to the presence of $c$, $V'(\rho_{c,0})$ is no longer equal to zero. Then, the Goldstone modes acquire a tiny mass $m_\varphi$ when $c$ is small. In this work, we  calculate $Z^{(t,i)}_{\varphi}$ and $m_\varphi$ using the fRG method. Note that the non-zero $m_{\varphi}$ means that the static correlator $G_{\varphi\varphi}^R(t\to 0^{+},\bm{x})$ contains a damping factor $\sim e^{-m_{\varphi}|\bm{x}|}$ in the real space. Following the QCD language, we call it the screening mass of pseudo-Goldstone modes. The dispersion relation can then be read off directly from the zero of the denominator in \Eq{eq:retardcorrelators}. We find a damped mode
\begin{align}
    \omega(q)=-\mathrm{i}\,\frac{Z^{(i)}_\varphi}{Z^{(t)}_\varphi}\,\left(m_\varphi^2+ q^2\right)\,, \label{eq:pseudodiffuisve}
\end{align}
which implies that the relaxation rate of the pseudo-Goldstone modes, that is the damping at zero wavenumber, reads $ \Omega_\varphi\equiv-\text{Im}\,\omega(q=0)$.
Note that \Eq{eq:pseudodiffuisve} is a rigid result that can be calculated fully numerically for all temperatures and all values of $m_\varphi$ (or $c$). 

In the pure SSB pattern, i.e., $m_{\varphi}\to 0$, we have that $\Omega\to 0$ and \Eq{eq:pseudodiffuisve} becomes a purely diffusive dispersion with a transport coefficient $D_\varphi\equiv Z^{(i)}_\varphi/Z^{(t)}_\varphi$ which is essentially the Einstein relation for the $\varphi$ fields, see Sec.~Methods for the details. We are more concerned with the effect from a non-zero $m_\varphi$ (or $c$). To analyze this explicitly, we start with the O($4$) model. Depending on whether the system is close to the critical temperature or not, the discussions will be separated as follows:

When $T$ is far below the critical temperature $T_c$ (but we still require that $T\gg m_\varphi$), the effect of the massive Higgs field $\delta \sigma$ can be ignored in the low energy description. In this region, one would expect that $Z_\varphi^{(t, i)}$ only receive corrections from the small parameter $m_\varphi/T$ in an analytic way. For this case, $\Omega_\varphi/m^2_{\varphi}$ should take the following expansion
\begin{align}
    \frac{\Omega_\varphi}{m^2_{\varphi}}\simeq D_{\varphi}(T)+\mathcal{O}\left(\frac{m_\varphi^2}{T^2}\right),\quad m_\varphi\ll T \ll T_c\,. \label{eq:analytic}
\end{align}
The analyticity of the expansion \Eq{eq:analytic} has been verified fully numerically in Sec.~Methods. Then, the deviation of  $\Omega_\varphi/m^2_{\varphi}$ from the value of $D_\varphi$ is parametrically suppressed for small $m_\varphi$ and the conventional damping-mass relation in \Eq{eq:universal} is obeyed.
   
When $T$ approaches the critical temperature from below, $T_c$ enters as a new scale. Throughout the work, $T_c$ is defined as the critical temperature in the purely SSB limit, i.e., when $c=0$. In this region, the system exhibits critical behaviors known as the static universality and critical slowing down which are insensitive to the microscopic physics but characterized by static critical exponents ($\alpha,\beta,\gamma,\delta,\nu,\eta$, only two of them are independent) as well as the dynamic critical exponent $z$, respectively \cite{Hohenberg:1977ym}. For convenience, we introduce a reduced temperature $t\equiv(T_c-T)/T_c$, which appears as an extra small parameter. Applying the scaling theory in the critical region, we derive that
\begin{align}
    \frac{Z^{(i)}_\varphi}{Z^{(t)}_\varphi}\propto m_{\varphi}^{(\eta_t-\eta)}\,,\quad m^2_{\varphi}\propto c^{\frac{2\nu}{\beta \delta} }\,\label{eq:Zratioc}
\end{align}
for $t/c^{1/(\beta\delta)}\ll 1$. The detailed analysis can be seen in Sec.~Methods. Since $m^2_\varphi$ does no longer depend on $c$ linearly, it violates the standard GMOR relation. This was already observed in the holographic AdS/QCD model close to the chiral phase transition \cite{Cao:2022csq}. Furthermore, we also show in Sec.~Methods that the mass of the Higgs mode shares the same scaling dependence on $c$ as the mass of the pseudo-Goldstones. Thus, the condition $m_\varphi\ll m_\sigma$ is no longer maintained and the Higgs mode should be considered as a light field in this region. 

Using the fRG method, we fix $\eta\approx 0.0374$ and $\eta_t\approx 0.0546$ in the O($4$) model. Combining \labelcref{eq:pseudodiffuisve} and \labelcref{eq:Zratioc}, one finds $\Delta_\eta\equiv \eta_t-\eta\approx0.0172\neq0$ implies the relation \labelcref{eq:universal} is not followed.

\begin{figure}[t]   \includegraphics[width=0.45\textwidth]{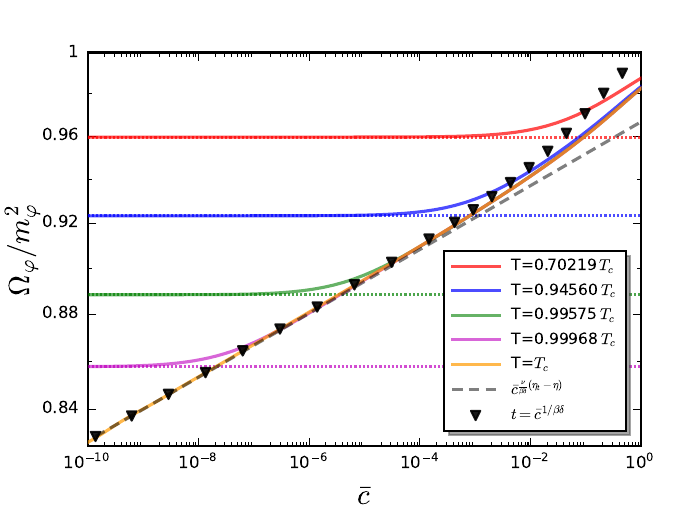}
  \caption{Ratio $\Omega_\varphi/m^2_\varphi$ as a function of $\bar{c}=c/c_0$ for various temperatures close to $T_c$, where $c_0$ denotes the value of $c$ with physical pion mass in QCD as shown in Sec.~Methods. The dotted lines correspond to values of $D_{\varphi}(T)$ in the pure SSB limit. The triangle points stand for results with a variational temperature with $t=\bar c^{1/{\beta \delta}}$. The gray dashed line denotes the scaling line $\sim \bar c^{\frac{\nu}{\beta \delta} (\eta_t-\eta)}$.}\label{fig:hydro_break}
\end{figure}

On the other hand, one can also read $\Omega_\varphi$ directly from the fully numeric result of \labelcref{eq:pseudodiffuisve}. It is found that the prediction above from the scaling theory perfectly matches the numeric result. In \Fig{fig:hydro_break}, we plot $\Omega_\varphi/m_\varphi^2$ as a function of the strength of the explicit symmetry breaking at several fixed values of temperature as well as a variational temperature with $t=\bar c^{1/{\beta \delta}}$, where $\bar{c}\equiv c/c_0$ is normalized with the value of $c$ corresponding to the physical pion mass in QCD, denoted by $c_0$, cf., the discussions in the following and in Sec.~Methods. For $\bar c \ll t^{\beta\delta}$, the ratio $\Omega_\varphi/m^2_\varphi$ approaches to the value of $D_\varphi$ as $c\rightarrow 0$. On the contrary, when $\bar c \gtrsim t^{\beta\delta}$, the anomalous scaling regime \labelcref{eq:scaling} takes over, resulting in a strong breakdown of relation \labelcref{eq:universal}. At the critical temperature $T=T_c$, the scaling regime stretches to $\bar{c}=0$, and therefore relation \labelcref{eq:universal} is not obeyed for any $c\neq 0$. In \Fig{fig:hydro_break} we also show the scaling line $\sim \bar c^{\frac{\nu}{\beta \delta} (\eta_t-\eta)}$, where the critical exponents are computed through fixed-point equations within the fRG approach. It is found that the numerical results at $T=T_c$ are in excellent agreement with the scaling line when  $c \to 0$, i.e., the pure SSB limit is approached.

\bmhead{Damping of pions near the chiral phase transition} 
The QCD matter below the critical temperature of chiral phase transition possesses a spontaneously broken approximate  $\text{SU(2)}_{\text{L}}\times \text{SU(2)}_{\text{R}}$ $\simeq$ O$(4)$ symmetry. Then, the critical dynamics of pions (the associated pseudo-Goldstone modes) is often investigated via the O($4$) model \cite{Son:1999pa,Son:2001ff, Son:2002ci,Grossi:2021gqi,Florio:2023kmy}. To that end, we adopt the identification, i.e., $\delta \sigma$ as the sigma meson and $\varphi\equiv\pi$ as pions.  A detailed map from the model parameters to the QCD observables for very low temperatures has been shown in Sec. Methods and TABLE \ref{tab:model-param}.

\begin{figure}[t]   
\includegraphics[width=0.45\textwidth]{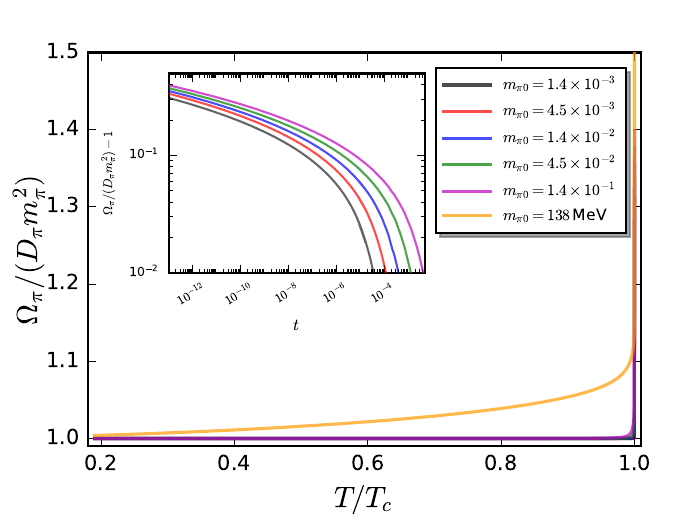}
  \caption{Ratio $\Omega_\pi/(D_\pi\,m_\pi^2)$ as a function of the temperature for different values of $m_{\pi 0}$, whose scaling behavior, i.e., the dependence of $\Omega_\pi/(D_\pi\,m_\pi^2)-1$ on the reduced temperature $t$ in a log-log plot, is analyzed in the inset.}\label{fig:ZvsT}
\end{figure}

\begin{figure}[t]
\includegraphics[width=0.45\textwidth]{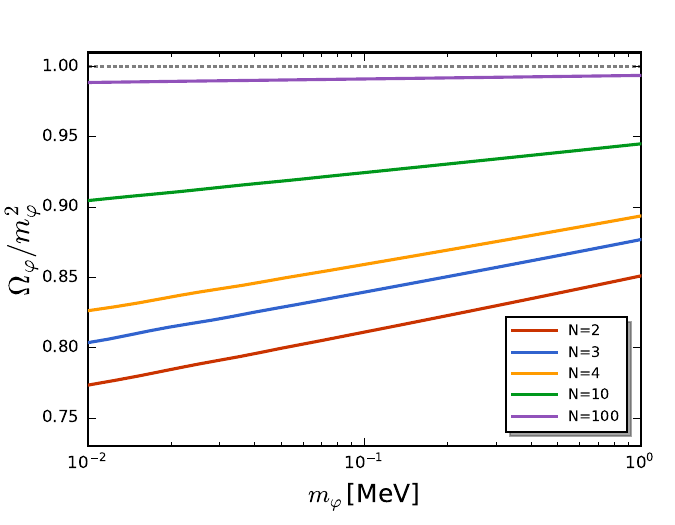}
  \caption{Ratio $\Omega_\varphi/m^2_\varphi$ as a function of $m_\varphi$ for different values of $N$ at $T=T_c$, where the gray dashed line denotes the value of $D_{\varphi}(T_c)$ in the large $N$ limit.}\label{fig:hydro_break_N}
\end{figure}

From the results of O($4$) model in \Fig{fig:hydro_break}, one finds that a deviation between the numerical results at $T=T_c$ and the scaling line takes place when $\bar{c} \gtrsim 10^{-6}\sim 10^{-4}$, corresponding to the pion mass at very low temperature $m_{\pi 0} \gtrsim 0.1 \sim 1$ MeV, from which one can estimate the size of the dynamic critical region, that is $m_{\pi 0} \lesssim 0.1 \sim 1$ MeV. Here, we use an extra subscript `0' to denote values of the observables at very low temperatures. In \Fig{fig:ZvsT}, we also plot $\Omega_\pi/(D_\pi\,m_\pi^2)$ as a function of $T/T_c$ for different values of $m_{\pi 0}$, i.e., different values of $c$ or $\bar{c}$ in \Fig{fig:hydro_break}. The ratio is very close to the unit at all temperatures far below $T_c$. However, its value becomes very sensitive to the value of $m_{\pi 0}$ near the critical point when $t\ll 1$. For the experiment result $m_{\pi 0}\approx 138$ MeV \cite{ParticleDataGroup:2020ssz}, we expect that the deviation from 1 becomes manifest ($\sim10\%$) at $T \approx 0.994\, T_c$. When $T\to T_c$, we find that the deviation displays a scaling growth
\begin{align}
    \frac{\Omega_\pi}{D_\pi m_\pi^2}-1\propto t^{-(\eta_t-\eta)\nu}, \label{tscaling2}
\end{align}
leading to $\Omega_\pi\gg D_\pi m_\pi^2$ close to the chiral phase transition. Note that this can never be observed in the AdS/QCD model where the large $N$ limit is taken \cite{Cao:2022csq}. 

Nevertheless, in reality, there exist couplings between the pseudo-Goldstone fields and the non-Abelian densities which lead to the propagation of pions. In the Supplementary Information, we have extended the studies from Model A to Model G in the Hohenberg-Halperin classification \cite{Hohenberg:1977ym}, and investigated both the damping and propagation of the pseudo-Goldstones. Note that the dynamic universality class for the chiral phase transition in two-flavor QCD can be described by Model G with $N=4$ \cite{Rajagopal:1992qz, Roth:2024rbi}. It is found that near the phase transition, the propagation of the pseudo-Goldstone slows down and coordinates with the damping, and their characteristic frequencies have a similar universal scaling behavior as \Eq{eq:scaling}  with $\Delta_\eta=-1/2$ which is no longer a small correction. Since the soft pions play the role of the leading non-hydrodynamic modes, these results will be featured in the real-time evolution of chiral observables as well as in the transport coefficients of QCD close to the critical point.

\bmhead{From finite N's to the large N limit} 
The O($N$) critical model can be applied to describe a wide range of interacting systems, including the Ising model ($N=1$), the XY model ($N=2$), the Heisenberg antiferromagnet ($N=3$) and the QCD ($N=4$) matter, etc. 

We now perform the fRG calculations for different values of $N$. The relevant results of the ratio $\Omega/m_\varphi^2$ as a function of the pseudo-Goldstone mass at $T=T_c$ are shown in \Fig{fig:hydro_break_N}. The tendency is that the scaling exponent $\Delta_\eta$ becomes smaller for larger $N$. In the limit of large $N$, one is able to calculate $\eta$ and $\eta_t$ analytically which are detailed in Sec.~Methods. The anomalous dimensions both approach zero in the way that $\{\eta,\,\eta_t\}\sim N^{-1}$ when $N\to \infty$. Despite that the system still experiences critical slowing down, the non-equilibrium fluctuations become negligible and $\Omega/m_\varphi^2$ is $c$-independent in the large $N$ limit. Moreover, we find that its value approaches to the value of $D_\varphi$. In this case, \Eq{eq:universal} is valid in the whole broken phase (as long as $c$ is small). Therefore, one should not expect that the novel scaling regime in \Eq{eq:scaling} can be observed in `mean-field systems' or holographic systems with classical gravity duals where the dynamic critical exponent is $z=2+\eta_t-\eta\rightarrow 2$ in the large $N$ limit \cite{Maeda:2009wv,Natsuume:2010bs}.

\section{Discussion}\label{sec3}
In this work, we investigate the damping mechanism of pseudo-Goldstones in a relaxational O($N$) model by applying the SK formalism of the fRG, aiming to extend the scope of the study to the critical region. It is found that, the conventional damping-mass relation in \Eq{eq:universal} governed by the hydrodynamic regime is only valid away from the critical point. In the vicinity of the critical point where the Higgs field becomes a light mode and its dynamics hence cannot be separated from the low energy description, the conventional relation is not obeyed. Instead, the pseudo-Goldstone damping displays a new universal behavior in \Eq{eq:scaling} controlled by the critical slowing down as well as the non-equilibrium fluctuations. Moreover, we also show how the critical damping evolves into the conventional one by increasing the value of $N$. Since our calculations are based on the effective action on the closed time contour, to the best of our knowledge, this is also the first time where the damping-mass relation of pseudo-Goldstones in a strongly coupled system has ever been  addressed at the stochastic level.

Our findings have the potential influence to some topics of high interest. For instance, the damping of the pseudo-Goldstone modes is closely related to that of the sigma mode which plays a pivotal role in determining the properties of critical slowing down near the critical end point (CEP) of QCD \cite{Arslandok:2023utm}. Thus, our findings are useful and valuable for the search for CEP, that is underway in heavy-ion collision experiments at many facilities around the world, e.g., Relativistic Heavy Ion Collider (RHIC) at Brookhaven. Related studies are in progress, and we will report it in the near future.
Since the spontaneous breaking of approximate symmetries abounds in both experiments and nature, our findings also have numerous applications in condensed matter physics, including pinned density waves/Wigner crystals \cite{Gruner:1994zz,Fogler:2000}, forced superfluid stripe phases \cite{Li:2021gpk}, frustrated magnets \cite{Tymoshenko2017,Jeffrey2018}, high $T_c$ superconductors \cite{Burgess:1998ku}, etc. The damping of the pseudo-Goldstone modes plays an important role in governing the large distance dynamics of these systems, for instance, the late time evolutions and the transport properties. 
Our results are significant for understanding the relevant effects in the critical region. It is expected that they can be tested in future experiments.

Furthermore, it should also be intriguing to verify our result using other methods that allow us to handle the critical fluctuations. For instance, one can include stochastic forces in the hydrodynamic equations following the line of \cite{Son:2002ci, Grossi:2020ezz, Grossi:2021gqi, Soloviev:2021syx, Florio:2021jlx, Florio:2023kmy} or calculate the `hydrodynamic loops' using the Schwinger-Keldysh field theory of fluctuating hydrodynamics \cite{Crossley:2015evo, Haehl:2015uoc, Liu:2018kfw, Chen-Lin:2018kfl, Jain:2020zhu, Jain:2020hcu, Donos:2023ibv}. Another direction is that one can enlarge the hydrodynamic description by considering the effects of Higgs mode in holographic systems near phase transitions \cite{Donos:2022xfd, Donos:2022qao}, and verify whether the conventional damping-mass relation is satisfied up to $T_c$ in the large $N$ limit.

\section{Methods}\label{sec4}
\bmhead{Setup of the relaxation model within the fRG} %\label{subsec2}
In the effective action in \Eq{eq:action}, the potential term on the Schwinger-Keldysh contour can be expressed as $V(\rho_+)-V(\rho_-)$ with
\begin{align}
    \rho_\pm\equiv\sum_{a=0}^{N-1}\frac{\phi^2_{a,\pm}}{2}\,,\label{eq:rhopm}
\end{align}
corresponding to the forward and backward time evolution branch, respectively. Taking the Keldysh rotation as follows
\begin{align}
    \phi_{a,\pm}=\frac{1}{\sqrt{2}}\left(\phi_{a,c}\pm\phi_{a,q}\right)\,,\label{eq:phipm}
\end{align}
and expanding the potential in terms of $q$-fields, we have 
\begin{align}
    V(\rho_+)-V(\rho_-)=V'(\rho_c)\,\phi_{a,q}\,\phi_{a,c}+\mathcal{O}(\phi_{a,q}^3)\,,\label{eq:Vpm}
\end{align}
up to the quadratic order.

From the real-time effective action in \Eq{eq:action}, one can obtain the flow equation of the effective potential within the fRG approach, which reads
\begin{align}
    \partial_\tau u'(\bar \rho)=&(-2+\eta)u'(\bar \rho)+(1+\eta)\bar \rho u^{(2)}(\bar \rho)\nonumber\\[2ex]
    &\hspace{-0.5cm}-\frac{1}{4\pi^2}\frac{2}{3}\left(1-\frac{\eta}{5}\right) \left[\frac{3u^{(2)}(\bar{\rho})+2\bar{\rho}u^{(3)}(\bar{\rho})}{\big(1+u'(\bar{\rho})+2\bar{\rho}u^{(2)}(\bar{\rho})\big)^2}\right.
    \nonumber\\[2ex]
    &\hspace{-0.5cm}+\left.(N-1)\frac{u^{(2)}(\bar{\rho})}{\big(1+u'(\bar{\rho})\big)^2}\right]\,,\label{eq:dtV1}
\end{align}
where we have defined the dimensionless, renormalized field and potential, viz.,
\begin{align}
    \bar \rho=Z^{(i)}_\varphi T^{-1} k^{-1} \rho\,,\qquad u(\bar \rho)=T^{-1} k^{-3}V(\rho)\,,\label{eq:barrho_u_def}
\end{align}
with the temperature $T$ and the RG scale $k$. Here we have used $\rho$ instead of $\rho_c$ for simplicity for the subscript of `classical'  field. The flow in \Eq{eq:dtV1} is formulated in terms of the RG time $\tau=\ln(k/\Lambda)$, where $\Lambda$ is some reference scale, e.g., an ultraviolet (UV) cutoff scale. Note that in order to arrive at the flow equation in \Eq{eq:dtV1} as well as those in what follows, we have used the optimized infrared (IR) regulator in fRG, see, e.g., \cite{Chen:2023tqc} for more details.

The static and dynamic anomalous dimensions are related to the spatial and temporal wave functions through
\begin{align}
    \eta=-\frac{\partial_\tau Z^{(i)}_{\varphi}}{Z^{(i)}_{\varphi}}\,,\qquad \eta_t=-\frac{\partial_\tau Z^{(t)}_{\varphi}}{Z^{(t)}_{\varphi}}\,,\label{eq:eta_etat_def}
\end{align}
respectively. Within the modified local potential approximation to the effective action in \Eq{eq:action}, usually called the $\mathrm{LPA}'$ truncation, one finds for the two anomalous dimensions
\begin{align}
    \eta&=\frac{2}{3\pi^2}\frac{\bar\rho_0 \big(u^{(2)}(\bar\rho_0)\big)^2}{\big(1+u'(\bar\rho_0)\big)^2 \big(1+u'(\bar\rho_0)+2\bar\rho_0 u^{(2)}(\bar\rho_0)\big)^2}\,,\label{eq:eta-expr}\\[2ex]
    \eta_t&=\frac{5-\eta}{30\pi^2}\bar \rho_0^2 \big(u^{(2)}(\bar\rho_0)\big)^2\nonumber\\[2ex]
    &\times
    \big(1+u'(\bar\rho_0)+2\bar\rho_0 u^{(2)}(\bar\rho_0)\big)^{-2}\nonumber\\[2ex]
    &\times\big(1+u'(\bar\rho_0)\big)^{-2}\big(1+u'(\bar\rho_0)+\bar\rho_0 u^{(2)}(\bar\rho_0)\big)^{-2}\nonumber\\[2ex]
    &\times\Big[4\big(1+u'(\bar\rho_0)\big)\big(1+u'(\bar\rho_0)+2\bar\rho_0 u^{(2)}(\bar\rho_0)\big)\nonumber\\[2ex]
    &+\big(1+u'(\bar\rho_0)\big)^2+\big(1+u'(\bar\rho_0)+2\bar\rho_0 u^{(2)}(\bar\rho_0)\big)^2\Big]\,,\label{eq:etat-expr}
\end{align}
where the potential is computed on the expectation value of the field, $\bar\rho_0$, which satisfies the equation of motion (EoM) as follows
\begin{align}
    \sqrt{2\rho_{0}}V'(\rho_{0})-c=0\,,\label{eq:EoM-rhoc}
\end{align}
with $\bar \rho_0=Z^{(i)}_\varphi T^{-1} k^{-1} \rho_{0}$. The EoM in \Eq{eq:EoM-rhoc} results from the effective action in \Eq{eq:action} by differentiating with respect to the `quantum' field.

In our numerical calculations, the flow equation of effective potential in \Eq{eq:dtV1} is integrated on a grid of the potential from a UV initial value $\Lambda$ to the IR limit $k \to 0$, with the static anomalous dimension in \Eq{eq:eta-expr} computed at each value of the RG scale. At the initial UV cutoff $k=\Lambda$, the effective potential is parameterized as 
\begin{align}
    V'_{k=\Lambda}(\rho)=\lambda_\Lambda(\rho-\rho_\Lambda)\,.\label{eq:V-UV}
\end{align}
Note only the derivative of potential is relevant, that is the reason why $V'$ rather than $V$ is specified in \Eq{eq:V-UV}. The initial UV scale $\Lambda$, the parameters $\rho_\Lambda$ and $\lambda_\Lambda$ in the potential at the initial scale, and the strength of explicit symmetry breaking $c$ as shown in the effective action in \Eq{eq:action} constitute the set of parameters for the relaxation model within the fRG. The values of model parameters adopted in this work are collected in TABLE \ref{tab:model-param}, where the observables produced at low temperature, say $T=1$ MeV, are also presented. In TABLE \ref{tab:model-param} the mass of pseudo-Goldstones, i.e., the pion mass in low energy QCD, $m_\varphi \equiv m_\pi =137.2$ MeV, is the physical mass in the real world. $m_\sigma$ denotes the mass of sigma mode. The pion decay constant at physical pion mass is given by $f_\pi=\sqrt{2\rho_{c0}}$, and $f_\pi^{\chi}$ stands for the pion decay constant in the chiral limit, to wit, vanishing pion mass. Moreover, within the parameters in TABLE \ref{tab:model-param} one finds the critical temperature of phase transition in the chiral limit $T_c=141.633955(10)$ MeV. This precision of $T_c$ is necessary for the scaling analysis in this work. Note that the existence of the anomalous damping of pseudo-Goldstone modes in the critical region found above, does not depend on the specific values of model parameters.

\begin{table*}[t]
  \caption{Observables and parameters in the relaxational O(4) model within the functional renormalization group approach.}\label{tab:model-param}
  \centering
  \begin{tabular}{c|c|c}
    \hline\hline & &    \\
    Observables \quad & \quad Values [MeV] \quad & \quad Parameters \\[2ex]
    \hline & & \\ [-1ex]
   $f_\pi$ & 93.5 & \quad $\Lambda=430\,\mathrm{MeV}$  \\[1ex]
   $m_\pi$ & 137.2 & \quad $\rho_\Lambda=3.84\times 10^3\,\mathrm{MeV}^2$  \\[1ex]
   $m_\sigma$ & 567.2 & \quad $\lambda_\Lambda=35$  \\[1ex]
   $f_\pi^{\chi}$ & 87.4 & \quad $c=1.76 \times 10^6\, \mathrm{MeV}^3$  \\[0ex]
   \hline\hline 
 \end{tabular}
\end{table*}

The values of critical exponents in the $\mathrm{LPA}^\prime$ truncation are found to be $\eta_t=0.054424$, $\eta=0.037320$, $\nu=0.78110$, $\beta=0.40513$, $\delta=4.7841$, which are calculated by solving the fixed-point equation, i.e., $\partial_\tau u'=0$ in \Eq{eq:dtV1}, see, e.g., \cite{Tan:2022ksv} for more details.

% The relevant data and codes are available in a public repository \cite{Tan2025}.
The relevant data is available in a Source Data File \cite{Tan2025}.

\bmhead{The Einstein relation for Goldstone fields in the pure SSB limit}
When the broken O($N$) symmetry is exact, i.e., $m_{\varphi}\to 0$, we have that $\Omega_\varphi\to 0$ and \Eq{eq:pseudodiffuisve} becomes a standard diffusive dispersion with the transport coefficient
\begin{align}
    D_\varphi(T)\equiv\frac{Z_\varphi^{(i)}(T,c=0)}{Z_\varphi^{(t)}(T,c=0)}\,. \label{eq:Gdiffusivity}
\end{align}
In this case, the physical meaning of $Z^{(t,i)}_{\varphi}(T,0)$ is manifest. From \Eq{eq:retardcorrelators}, we have that
\begin{align}
G_{\varphi\varphi}^R(\omega,q)=\frac{\mathrm{i} Z^{(t)}_\varphi(T,0) \omega+Z^{(i)}_\varphi(T,0) q^2}{\big(Z^{(t)}_\varphi(T,0) \omega \big)^2+\big(Z^{(i)}_\varphi(T,0) q^2 \big)^2}\,.\label{eq:retardcorrelatorszeroc}
\end{align}
Then, $1/Z^{(t)}_\varphi(T,0)$ plays the role of the `Goldstone conductivity' satisfying the Kubo formula for the Goldstone fields
\begin{align}
    \sigma_\varphi\equiv\frac{1}{Z^{(t)}_\varphi(T,0)}=\lim_{\omega\rightarrow 0}\,\omega\, \text{Im}\,G_{\varphi\varphi}^R(\omega,q=0)\,, \label{eq:kubo}
\end{align}
Note that \Eq{eq:kubo} can also be recast as the standard Kubo formula for the retarded correlator of $\partial_t \varphi$ like in \cite{Amoretti:2018tzw,Amoretti:2019cef}. Furthermore, defining $\lambda=\partial _i\varphi$, it is obvious to see that 
\begin{align}
    \chi_{\lambda\lambda}\equiv\frac{1}{Z^{(i)}_\varphi(T,0)}=\lim_{q\rightarrow 0} q^2\,G_{\varphi\varphi}^R(0,q)\,. \label{eq:chi}
\end{align}
Therefore, the inverse of the Goldstone stiffness $1/Z^{(i)}_{\varphi}(T,0)$ can be understood as the `Goldstone susceptibility' \cite{Cao:2022csq}. Then, \Eq{eq:Gdiffusivity} is essentially an Einstein relation for the Goldstone fields in the pure SSB pattern.

\begin{figure}[t]
  \includegraphics[width=0.45\textwidth]{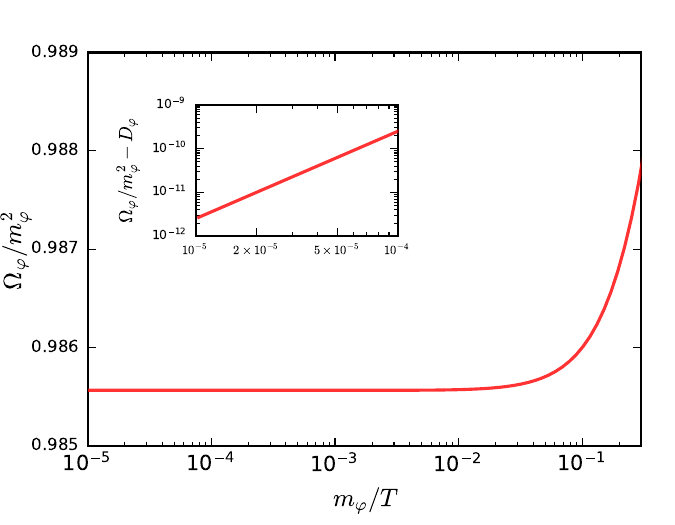}
  \caption{Ratio $\Omega_\varphi/m^2_\varphi$ as a function of $m_\varphi/T$ with $T=0.3\,T_c$ and $N=4$. The inset shows the subleading term of mass correction, i.e., $\Omega_\varphi/m_\varphi^2-D_\varphi$ as a function of $m_\varphi/T$.}\label{fig:subleading}
\end{figure}

\bmhead{Numerical verification of \Eq{eq:analytic}}
In \Fig{fig:subleading} we show the dependence of $\Omega_\varphi/m^2_\varphi$ on the mass of pseudo-Goldstones, where the temperature is chosen a value, $T=0.3\,T_c$ far lower than the critical temperature. The subleading term of the expansion in \Eq{eq:analytic} is shown in the inlay, from which one can extract the exponent $\alpha$ via
\begin{align}
    \frac{\Omega_\varphi}{m_\varphi^2}-D_\varphi(T) \propto \left(\frac{m_\varphi}{T}\right)^{\alpha}\,.\label{checkddrelation}
\end{align}
Our result shows that $\alpha=2.00021(2)$ which is equal to 2 within errors.

\bmhead{Scaling analyses of the wave functions, pseudo-Goldstone and Higgs masses}
In this section, we perform scaling analyses for the wave functions, pseudo-Goldstone and Higgs masses. Note that these analyses are general and independent of the computational method we use, e.g., the fRG. In the critical region, the two wave functions in \Eq{eq:action} read
\begin{align}
    Z^{(i)}_\varphi=t^{-\nu \eta}f^{(i)}(\mathbb{z})\,,\quad Z^{(t)}_\varphi=t^{-\nu \eta_t}f^{(t)}(\mathbb{z})\,, \label{eq:scaling-func}
\end{align}
with a scaling variable $\mathbb{z}\equiv t c^{-1/(\beta\delta)}$, where $f^{(i)}(\mathbb{z})$ and $f^{(t)}(\mathbb{z})$ are universal scaling functions. In the case of $c \to 0$, the two scaling functions approach towards constants, which leave us with 
\begin{align}
    \frac{Z^{(i)}_\varphi}{Z^{(t)}_\varphi}\propto t^{\nu (\eta_t-\eta)}\,. \label{tscaling}
\end{align}
In the other case of $t \to 0$, which is our most concern in this work, the variable $t$ in \Eq{eq:scaling-func} has to be eliminated by virtue of the scaling functions. Then, we arrive at 
\begin{align}
    \frac{Z^{(i)}_\varphi}{Z^{(t)}_\varphi}\propto c^{\frac{\nu}{\beta \delta} (\eta_t-\eta)}\,.\label{eq:Zratioc2}
\end{align}
On the other hand, employing the EoM of fields in \Eq{eq:EoM-rhoc}, one arrives at the mass of the pseudo-Goldstone modes immediately as
\begin{align}   m_{\varphi}^2=\frac{V'(\rho_{0})}{Z^{(i)}_{\varphi}}=\frac{c}{\sigma_0 Z^{(i)}_{\varphi}}\,,\label{eq:mvarphi2-scaling-Sup0}
\end{align}
where the order parameter $\sigma_0=\sqrt{2\rho_{0}}$ plays the role of magnetization. When the reduced temperature $t \to 0$ and in the critical region, one has $\sigma_0 \sim c^{1/\delta}$. Moreover, one finds from the scaling function of the spatial wave function in \Eq{eq:scaling-func} $Z^{(i)}_{\varphi} \sim c^{-\nu \eta/(\beta \delta)}$. Consequently, in the case of $t \to 0$ one is led to
\begin{align}
    m_{\varphi}^2\propto \frac{c}{c^{\frac{1}{\delta}} c^{-\frac{\nu \eta}{\beta \delta}}}=c^{\frac{2\nu}{\beta \delta}}\,.\label{eq:mvarphi2-scaling-Sup}
\end{align}
In the last step of \Eq{eq:mvarphi2-scaling-Sup} we have used the scaling relations for different critical exponents as follows
\begin{align}
    \beta = \frac{\nu}{2}(d-2+\eta)\,,\qquad \delta = \frac{d+2-\eta}{d-2+\eta}\,,\label{eq:beta_delta}
\end{align}
where $d$ denotes the spatial dimension. Combining \Eq{eq:Zratioc2} and \Eq{eq:mvarphi2-scaling-Sup} one finds
\begin{align}
    \frac{Z^{(i)}_\varphi}{Z^{(t)}_\varphi}\propto m_{\varphi}^{(\eta_t-\eta)}\,,\label{eq:Zratioc3}
\end{align}
which is the relation in \Eq{eq:Zratioc}.
 
We proceed with the scaling analysis of the mass of sigma mode. It is well known that the sigma mass is related to the correlation length $\xi$ through $m_\sigma=\xi^{-1}$, and the correlation length behaves as $\xi \sim t^{-\nu}$. Thus, one finds the scaling behavior of the sigma mass
\begin{align}
    m_\sigma^2=t^{2\nu}f^{\sigma}(\mathbb{z})\,,\label{eq:scaling-func-msig}
\end{align}
with the scaling function $f^{\sigma}(\mathbb{z})$. In the case of $t \to 0$, the reduced temperature in \Eq{eq:scaling-func-msig} has to be eliminated by the scaling function, which leaves us with
\begin{align}
    m_\sigma^2 \sim c^{\frac{2\nu}{\beta \delta}}\,.\label{eq:msig2-scaling}
\end{align}
Comparing \Eq{eq:mvarphi2-scaling-Sup} and \Eq{eq:msig2-scaling}, we find that both the Goldstone and sigma masses have the same scaling dependence on the strength of symmetry breaking $c$ in the critical region.

\begin{figure}[t]
\includegraphics[width=0.45\textwidth]{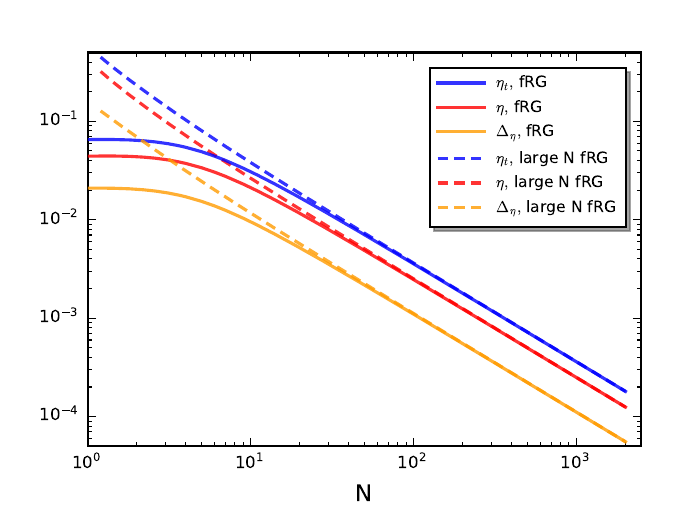}
  \caption{Dynamic anomalous dimension $\eta_t$, static anomalous dimension $\eta$ and their difference $\Delta_\eta\equiv\eta_t-\eta$ as functions of $N$. Here, the solid lines denote the full numerical results obtained in \labelcref{eq:dtV1,eq:eta-expr,eq:etat-expr}, and the dashed ones denote those obtained from the expressions of large $N$ in \Eq{eq:eta-largeN}.}\label{fig:Large_N_eta}
\end{figure}

\bmhead{Anomalous dimensions in the large $N$ limit}
Now, we discuss the large $N$ model. In the large $N$ limit, the flow of effective potential in \Eq{eq:dtV1} is simplified as
\begin{align}
    &\partial_\tau u'(\bar \rho)\nonumber\\[2ex]
    =&(-2+\eta)u'(\bar \rho)+(1+\eta)\bar \rho u^{(2)}(\bar \rho)\nonumber\\[2ex]
    &-\frac{1}{4\pi^2}\frac{2}{3}\left(1-\frac{\eta}{5}\right)(N-1)\frac{u^{(2)}(\bar{\rho})}{\big(1+u'(\bar{\rho})\big)^2}\,.\label{eq:dtV1-largeN}
\end{align}
We are interested in the fixed-point solution of the flow equation above, i.e., $\partial_\tau u'_{*}(\bar \rho)=0$. The fixed-point equation in \Eq{eq:dtV1-largeN} can be solved analytically through an implicit formulation, whose solution reads
\begin{align}
    \bar\rho=&\frac{1}{30\pi^2}\, {}_2F_1\left(2,1+\frac{3}{\eta-2};\frac{2\eta-1}{\eta-2};-u'_{*}(\bar\rho)\right)\nonumber\\[2ex]
    &\times\frac{(5-\eta)}{1+\eta}(N-1)+\mathcal{C}  u'_{*}(\bar \rho)^{\frac{1+\eta}{2-\eta}}\,,\label{eq:solu-fixpoin}
\end{align}
with a constant $\mathcal{C}$ to be determined, where ${}_2F_1$ is a hypergeometric function with a general notation ${ }_p F_q\left(a_1, \ldots, a_p ; b_1, \ldots, b_q ; z\right)$. As for the last term in \Eq{eq:solu-fixpoin}, since the exponent $(1+\eta)/(2-\eta)$ is not an integer in general, which would result in a branch cut for $u'_{*}(\bar\rho)< 0$. In consequence, the constant $\mathcal{C}$ has to be vanishing to make sure that the potential is analytic. The minimal point of the potential is determined by the EoS in \Eq{eq:EoM-rhoc}. In the chiral limit one has $u'_{*}(\bar\rho_0)=0$, which leaves us with
\begin{align}
    \bar\rho_{0}&=\frac{1}{30\pi^2}\frac{(5-\eta)}{(1+\eta)}(N-1)\,,\nonumber\\[2ex] 
    u^{(2)}_{*}(\bar\rho_0)&=15\pi^2\frac{(1-2\eta)}{(5-\eta)}\frac{1}{(N-1)}\,. \label{eq:rho0_upp_largeN}
\end{align}
Inserting the expressions above as well as $u'_{*}(\bar\rho_0)=0$ into \Eq{eq:eta-expr} and \Eq{eq:etat-expr}, one obtains analytic expressions of the static and dynamic anomalous dimensions as follows
\begin{align}
    &\eta=\frac{5}{N-1}\frac{(1+\eta)(1-2\eta)^2}{(5-\eta)(2-\eta)^2}\,,\nonumber\\[2ex]
    &\eta_t=\frac{1}{9(N-1)}\frac{(1-2\eta)^2(13+15\eta-2\eta^3)}{(2-\eta)^2}\,.\label{eq:eta-largeN}
\end{align}
These are implicit functions. But one can still see clearly that $\eta \to 0$ and $\eta_t \to 0$ when $N \to \infty$. Then, the scaling exponent $\Delta_\eta\equiv \eta_t-\eta$ also becomes trivial in the large $N$ limit. Numerical calculations of \Eq{eq:eta-largeN} are presented in \Fig{fig:Large_N_eta} in dashed lines, which are also compared with the full results obtained in \labelcref{eq:dtV1,eq:eta-expr,eq:etat-expr}.

In the following we adopt the method of eigenperturbations to calculate the critical exponent $\nu$ in the large $N$ limit, cf. e.g., \cite{Tan:2022ksv}. A small perturbation of the effective potential around the fixed point can be described as
\begin{align}
    u(\bar\rho)=u_{*}(\bar\rho)+\epsilon\,\mathrm{e}^{-\omega \tau}v(\bar\rho)\,,\label{eq:eigenpertur}
\end{align}
with a small parameter $\epsilon$, where $\omega$ and $v(\bar\rho)$ stand for the eigenvalue and eigenfunction of the perturbation, respectively. Substituting \Eq{eq:eigenpertur} into \Eq{eq:dtV1-largeN}, one is led to
\begin{align}
    \omega v(\bar\rho)=&3 v(\bar\rho)-(1+\eta)\bar\rho\, v'(\bar\rho)\nonumber\\*[2ex]
    &+\frac{5-\eta}{30\pi^2}\frac{(N-1)v'(\bar\rho)}{\big(1+u_{*}^{\prime}(\bar\rho)\big)^2}\,.\label{eq:dvdrho}
\end{align}
This equation can be solved analytically, and its solution is given by
\begin{align}
    v(\bar\rho)=\mathcal{C}_v \Big [u'_{*}(\bar\rho)\Big ]^{\frac{3-\omega}{2-\eta}}\,,\label{eq:eigenfun-largeN}
\end{align}
with a constant $\mathcal{C}_v$. Since the Wilson-Fisher fixed point has the property $u'_{*}(\bar\rho=0)<0$, in order to warrant that the eigenfunction in \Eq{eq:eigenfun-largeN} is analytic, one has to choose the exponent there to be positive integers, i.e.,
\begin{align}
    \frac{3-\omega_n}{2-\eta}=n+1\,, \qquad n=0,1,2,\cdots\,.\label{eq:omegan}
\end{align}
Thus, we are led to 
\begin{align}
    \omega_n=3-(n+1)(2-\eta)\,. \label{eq:omegan_sol}
\end{align}
As we have shown above, $\eta \to 0$ when $N \to \infty$. So, only the first eigenvalue is relevant, i.e., $\omega_0=1+\eta>0$. The critical exponent reads
\begin{align}
    \nu=\frac{1}{\omega_0}=\frac{1}{1+\eta}\,. \label{eq:nu_largeN}
\end{align}

\begin{figure}[t]
  \includegraphics[width=0.45\textwidth]{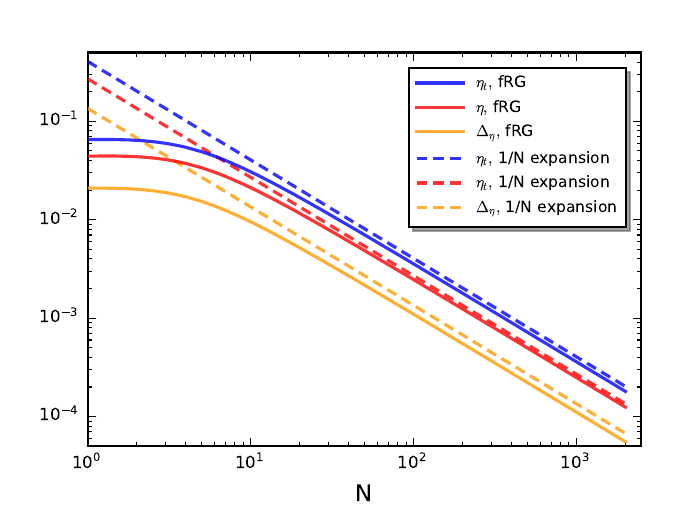}
  \caption{Anomalous dimensions $\eta_t$, $\eta$ and their difference $\Delta_\eta\equiv\eta_t-\eta$ as functions of $N$ with spatial dimension $d=3$ obtained in fRG with $\mathrm{LPA}'$ truncation, in comparison to the results of $1/N$ expansion in the order of $N^{-1}$ \cite{Ma:1973cmn, Halperin:1972bwo}.}\label{fig:eta-invN}
\end{figure}

\bmhead{Comparison to the $1/N$ expansion and $\epsilon=4-d$ expansion}
In this section, we compare our results with those of $1/N$ expansion and $\epsilon=4-d$ expansion. The static anomalous dimension is calculated up to the order of $N^{-1}$ in the $1/N$ expansion \cite{Ma:1973cmn}, as follows
\begin{align}
    \eta&=4 N^{-1}\Big(\frac{4}{d}-1\Big) S_d+\mathcal{O}\left(N^{-2}\right)\,,\label{eq:eta-invN-expan}
\end{align}
with
\begin{align}
    S_d=\frac{\sin\bigg[ \pi\Big(\frac{d}{2} -1\Big)\bigg]}{\pi\Big(\frac{d}{2} -1\Big)B\Big(\frac{d}{2} -1, \frac{d}{2} -1\Big)}\,, 
\end{align}
where $B$ is the beta function. The relevant dynamic anomalous dimension reads \cite{Halperin:1972bwo}
\begin{align}
    &\eta_t\nonumber\\[2ex]
    =&\Bigg[\Big(\frac{4}{4-d}\Big)\bigg(\frac{d B\left(\frac{d}{2}-1, \frac{d}{2} -1\right)}{8 \int_0^{1 / 2} d x[x(2-x)]^{d / 2-2}}-1\bigg)+1\Bigg]\eta.
\label{eq:etat-invN-expan}
\end{align}
In \Fig{fig:eta-invN} we show the dynamic, static anomalous dimensions and their difference obtained in fRG with $\mathrm{LPA}'$ truncation, in comparison to the results of $1/N$ expansion in the order of $N^{-1}$. One can see that the results are consistent.

\begin{figure}[t]
  \includegraphics[width=0.45\textwidth]{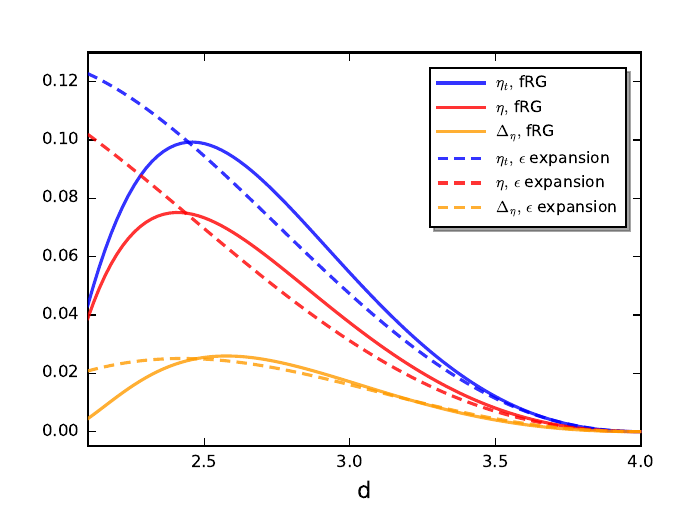}
  \caption{Anomalous dimensions $\eta_t$, $\eta$ and their difference $\Delta_\eta\equiv\eta_t-\eta$ as functions of the spatial dimension $d$ for the $O(4)$ symmetry obtained in fRG with $\mathrm{LPA}'$ truncation, in comparison to the results of $\epsilon$ expansion in the order of $\epsilon^4$ \cite{Wilson:1973jj, Halperin:1972bwo, Antonov:1984yva, Adzhemyan:2017jcv}.}\label{fig:eta-epsilon}
\end{figure}

As for the $\epsilon=4-d$ expansion, the static anomalous dimension reads \cite{Wilson:1973jj}
\begin{align}
    &\eta\nonumber\\[2ex]
    =&\frac{N+2}{2(N+8)^2}\epsilon^2+\frac{N+2}{2(N+8)^2}\left[\frac{6(3N+14)}{(N+8)^2}-\frac{1}{4}\right]\epsilon^3\nonumber\\[2ex]
    &+\frac{N+2}{32(N+8)^6}\Big[-5N^4-230N^3+1124N^2 \nonumber\\[2ex]
    &-768(5N+22)(N+8)\times 0.60103 \nonumber\\[2ex]
    &+17920N+46144\Big]\epsilon^4+\mathcal{O}(\epsilon^5)\,,\label{eq:eta-epsilon}
\end{align}
that is expanded up to the order of $\epsilon^4$. The dynamic anomalous dimension obtained in the $\epsilon$ expansion \cite{Halperin:1972bwo, Antonov:1984yva, Adzhemyan:2017jcv} is given by
\begin{align}
	\eta_t=&\bigg[1+\Big(1-0.188483417\epsilon-0.099952926\epsilon^2\nonumber\\[2ex]&+\mathcal{O}(\epsilon^3)\Big)
    \times \bigg(6\ln\Big(\frac{4}{3}\Big)-1\bigg) \bigg]\eta\,.\label{eq:etat-epsilon}
\end{align}
In \Fig{fig:eta-epsilon} we compare the $\epsilon$ expansion with the fRG. It is found that these two calculations are in qualitative agreement with each other when $d \gtrsim 2.5$. The anomalous dimension should be vanishing in the limit $d \to 2$ when $N \geq 2$ based on the Mermin-Wagner-Hohenberg theorem \cite{Mermin:1966fe, Hohenberg:1967zz, Coleman:1973ci}, and apparently the $\epsilon$ expansion is not reliable any more in low dimension.

\backmatter

\bmhead{Data availability}
The source data underlying all figures are available at \url{https://doi.org/10.6084/m9.figshare.28329866}.

\bmhead{Code availability}
The code used to solve the fRG equations is publicly available at \url{https://doi.org/10.6084/m9.figshare.28329866}.

\bmhead{Acknowledgements}
We would like to thank Matteo Baggioli, Blaise Gout\'eraux, Jan M. Pawlowski, Fabian Rennecke, Nicolas Wink and Hongbao Zhang for inspiring discussions and reading a previous version of this manuscript. W.J.F. is supported by the National Natural Science Foundation of China under Grant Nos.\ 12447102, 12175030 and 12147101. W.-J.L. is supported by the National Natural Science Foundation of China under Grant Nos.\ 12275038 and 12047503. W.-J.L. also acknowledges the sponsorship from the Peng Huanwu Visiting Professor Program in 2023 and Institute of Theoretical Physics, Chinese Academy of Sciences for the warm hospitality during which part of this work was completed.

\bmhead{Author contributions}
Y.-y.T. and Y.-r.C. performed the numerical computation and drew the figures; W.-j.F. and W.-J.L. analyzed the results and supervised the whole project; W.-j.F. and W.-J.L. cowrote the manuscript, and all authors contributed the discussions of the results and development of the manuscript.

\bmhead{Competing interests}
The authors declare no competing interests.

\clearpage
\includepdf[pages=-]{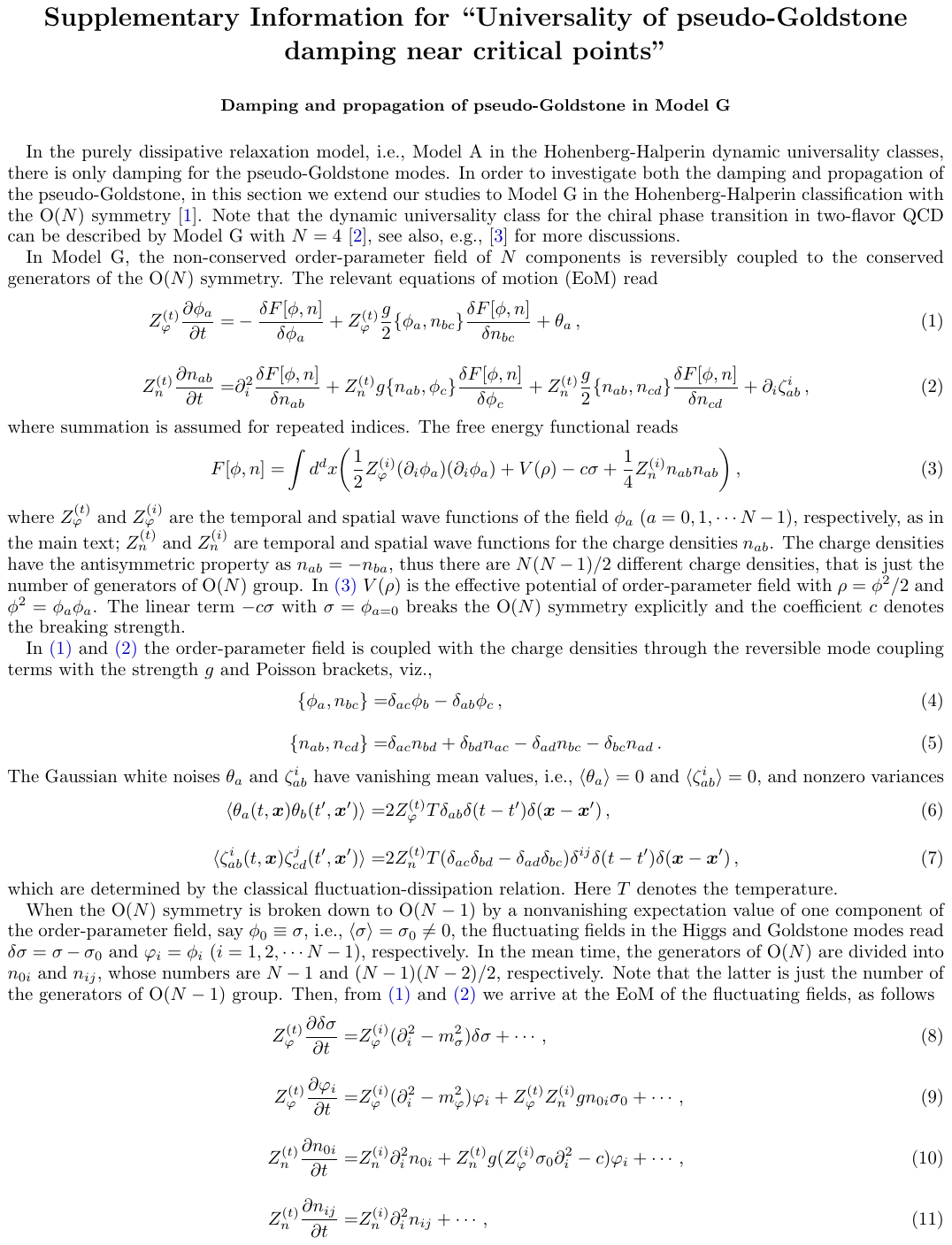}

\end{document}